\documentclass[12pt]{article}
\newcommand{\be}{\begin{equation}}
\newcommand{\ee}{\end{equation}}
\newcommand{\bt}{\tilde{b}}
\newcommand{\At}{\tilde{A}}
\newcommand{\bi}{\begin{itemize}}
\newcommand{\ei}{\end{itemize}}
\newcommand{\bea}{\begin{eqnarray}}
\newcommand{\eea}{\end{eqnarray}}
\newcommand{\ba}{\begin{array}}
\newcommand{\ea}{\end{array}}

\newcommand{\nn}{\nonumber}
\usepackage[left=2.50cm, right=2.50cm, top=2.50cm, bottom=2.50cm]{geometry}
\usepackage[utf8]{inputenc}
\usepackage{cancel}
\usepackage[english]{babel}
\usepackage{physics}
\usepackage{hyperref}
\usepackage{amsthm}
\usepackage{mathtools}
\usepackage{graphicx}
\usepackage{ulem}
\usepackage{changepage}
\usepackage{amssymb,amsmath}
\usepackage{verbatim}
\usepackage{empheq}
\usepackage{xcolor}
\usepackage{tikz}
\usepackage{float}
\usepackage{multirow}
\usepackage{multicol}
\usepackage{amsmath}
\usepackage{centernot}
\usepackage[hang,flushmargin]{footmisc} 
\usetikzlibrary{tikzmark}
\numberwithin{equation}{section}
\hypersetup{hidelinks}

\newlength{\bibitemsep}
\newlength{\bibparskip}
\let\oldthebibliography\thebibliography
\renewcommand\thebibliography[1]{%
  \oldthebibliography{#1}%
  \setlength{\parskip}{\bibitemsep}%
  \setlength{\itemsep}{\bibparskip}%
}
\begin{document}
\par
\bigskip
\Large
\noindent
{\bf The theory of symmetric tensor field: from fractons to gravitons and back\
\bigskip
\par
\rm
\normalsize

\hrule

\vspace{1cm}

\large
\noindent
{\bf Alberto Blasi$^{1,a}$}, 
{\bf Nicola Maggiore$^{1,2,b}$}\\

\par
\small
\noindent$^1$ Dipartimento di Fisica, Universit\`a di Genova, Italy.
\smallskip

\noindent$^2$ Istituto Nazionale di Fisica Nucleare - Sezione di Genova, Italy.

\smallskip

\vspace{1cm}

\noindent
{\tt Abstract~:}

We consider the theory of a symmetric tensor field in 4D, invariant under a subclass of infinitesimal diffeomorphism transformations, where the vector diff parameter is the 4-divergence of a scalar parameter. The resulting gauge symmetry characterizes the ``fracton'' quasiparticles and identifies a theory which depends on a dimensionless parameter, which cannot be reabsorbed by a redefinition of the tensor field, despite the fact that the theory is free of interactions. This kind of ``electromagnetic  gauge symmetry'' is weaker that the original diffeomorphism invariance, in the sense that the most general action contains, but is not limited to, linearized gravity, and we show how it is possible to switch continuously from linearized gravity to a mixed phase where both gravitons and fractons are present, without changing the degrees of freedom of the theory. The gauge fixing procedure is particularly rich and rather peculiar, and leads to the computation of propagators which in the massive case we ask to be tachyonic-free, thus constraining the domain of the parameter of the theory. Finally, a closer contact to fractons is made by the introduction of a parameter related to the ``rate of propagation''. For a particular value of this parameter the theory does not propagate at all, and we guess that, for this reason, the resulting theory should be tightly related to the fracton excitations.

\vspace{.5cm}

\vspace{\fill}
\noindent{\tt Keywords:} 
Quantum Field Theory, Fractons, Linearized Gravity

\vspace{1cm}

\hrule
\noindent{\tt E-mail:
$^a$alberto.blasi@ge.infn.it,
$^b$nicola.maggiore@ge.infn.it.
}
\newpage

\section{Introduction}

Recently, quantum phases of matter have been studied which are characterized by fractionalized quasiparticles of the same type of spin liquids and fractional quantum Hall systems \cite{Prem:2017kxc,Pretko:2016lgv,Pretko:2017fbf,Pretko:2018jbi,Nandkishore:2018sel}, which are commonly described in terms of gauge theories \cite{Fradkin:1991wy,Zhang:1992eu,Cho:2010rk}. But, differently from the latter, these new 4D phases are described by higher rank symmetric tensor fields, rather than vectors \cite{Slagle:2018kqf,Pretko:2020cko,Seiberg:2020wsg,Argurio:2021opr,Jain:2021ibh}. The corresponding fractionalized quasiparticles have been called ``fractons'', which are excitations of $U(1)$ gauge field theories which generalize the usual electromagnetism. The defining property of these quasiparticles is their complete or partial immobility \cite{Chamon:2004lew,Vijay:2015mka,Vijay:2016phm}. This property may be achieved through an extension of the usual electromagnetic theory, where the spatial vector gauge field $A_i$ is substituted by a rank-2 spatial symmetric tensor gauge field $A_{ij}$. Now, the usual gauge symmetry 
\begin{equation}
\delta A_i=\partial_i\Phi
\label{1.1}\end{equation}
may be extended in two different ways \cite{Pretko:2017xar}~:
\begin{itemize}
\item by means of a {\bf vector charge theory}
characterized by an unusual vector charge density $\rho^i$ and by an extended gauge transformation of the gauge tensor $A_{ij}$ with a vector ghost field
\begin{equation}
\delta A_{ij}=\partial_i\Phi_j+\partial_j\Phi_i\ ,
\label{1.2}\end{equation}
which is the usual (spatial) infinitesimal diffeomorphism invariance, or

\item through a {\bf scalar charge theory}
characterized by the usual scalar charge density $\rho$ and by an unusual extended gauge transformation of the gauge tensor $A_{ij}$ with a scalar ghost field\ .
\begin{equation}
\delta A_{ij}=\partial_i\partial_j\Phi\ .
\label{1.3}\end{equation}
\end{itemize}
Both extensions imply restrictions on the mobility of the excitation, which is partial in the case of vector charge theory and complete in the scalar charge theory \cite{Pretko:2017xar}.
Fractons are commonly understood as belonging to the second type of realization of the higher rank gauge theory, namely  the scalar charge theory. This type of new quasiparticles lies on the boundary of several theoretical research activities, wherever immobility plays an important role (amorphousness in solids \cite{Chamon:2004lew,Prem:2017qcp, pretkoscreening} and in computational systems \cite{Haah:2011drr,Bravyi:quantum,Vijay:2017cti}, 
long-term entanglement \cite{Vijay:2015mka,Vijay:2016phm,Shi:2017qdx}, for instance). 
In this paper we would like to focus on the consequences of dealing with a gauge field $A_{ij}$ with a tensorial nature, which necessarily shares its birth certificate with Linearized Gravity (LG) \cite{Carroll.book}, as it is apparent from the two cases mentioned above. It is readily seen, indeed, that for a particular choice of the vector parameter of the spatial diffeomorphism transformation \eqref{1.2}
\begin{equation}
\Phi_i = (\partial_i\Phi)/2
\label{1.4}\end{equation}
the first transformation \eqref{1.2} turns into the second \eqref{1.3}. This suggests that a theory invariant under the transformation \eqref{1.2} should be invariant also under \eqref{1.3}, being a particular case, but not {\it vice versa}. Which means that LG should be a particular case of the more general fracton theory invariant under \eqref{1.3}. 

This fact has already been observed \cite{Pretko:2017fbf}  and in this paper, we aim at providing more details on this subject. The point of view that we take, where we consider a covariant 4D extension of the fracton transformation of a rank-2 symmetric tensor (given in
\eqref{1.3}), has not been explored in the literature. 
We want to study the resulting tensor gauge field theory, using standard Quantum Field Theory techniques. The starting point is the necessary gauge fixing procedure, which already displays some non-standard features, on which we will elaborate. We will then discuss in which sense fractons and gravitons are related, what commonalities they have, and how they differ. We will then consider the case where the fractons are massive, also exhibiting some unexpected features.
The first notable difference with LG is that we have at our disposal a dimensionless parameter $a$, and second that in order to find the propagators, the gauge fixing is much simpler. Nevertheless the propagator structure turns out to be quite rich and its study leads to a ``phase'' diagram of the theory which includes particular values of the free parameter $a$ where the invariant action acquires a different structure. 
In our analysis we will adopt the simplest possible Landau gauge which has the advantage of making the ``physics'' more transparent and the computations much easier. Particular attention has ben paid to the degrees of freedom (dof) of the theory as the $a$-parameter moves on the real line. We shall also see that the phase diagram suggests an interpretation of the $a$-parameter and a possible connection with fractons. At the end we will explore, as is usual in LG, the massive extension of the model and will find that, in order to avoid tachyonic poles, a segment of the real line must be excluded from the domain of $a$. The  paper is organised as follows. In Section 2 the invariant action for a 4D symmetric tensor field  is derived as the most general one invariant under the particular infinitesimal symmetry characterising the ``emergent electromagnetism'' described in \cite{Pretko:2016lgv,Pretko:2017fbf}. We then analyze the gauge fixing procedure, and we show that the Landau gauge choice is the natural one for this model. Finally, we compute the propagators, which lead to identify some critical values of the parameter $a$. One of these special values, notably $a=0$, is expected since the model acquires a stronger symmetry and hence a completely different structure and while another one is also expected where the symmetric tensor becomes traceless, there is a third intriguing singular point, $a=1$, where the model can be written in terms of a non propagating, invariant vector field. In Section 3 the degrees of freedom are studied, which allow for a further interpretation of the $a$-parameter, which turns out to be related also to the gauge fixing structure of the theory, $i.e.$ the choice of a scalar ghost. The massive case is considered in Section 4, where appear analogies and differences with respect to LG: as in the Fierz-Pauli theory of LG only one mass parameter is allowed instead of two \cite{Gambuti:2021meo,Gambuti:2020onb}, as a symmetric tensor field would in principle admit, and, differently to what happens in massive LG \cite{Hinterbichler:2011tt,deRham:2014zqa,Boulware:1972yco}, the number of degrees of freedom does not depend on the presence of a mass term. Finally, in Section 5 we discuss the main feature of fractons, $i.e.$ the fact of having limited or even zero mobility \cite{Chamon:2004lew,Vijay:2015mka,Vijay:2016phm}, and we propose a parameter, which of course is a function of $a$, which measures the rate of propagation of the particle described by this theory. We show that indeed for a particular value of this parameter the corresponding particle does not propagate, which lead us to identify the parameter, hence the theory, corresponding to fractons.

\section{The model}
\subsection{The symmetry and the invariant action}

The action is built in flat 4D spacetime with a symmetric rank-2 tensor field $A_{\mu\nu}(x)$, and is invariant under the gauge transformation \eqref{fractonsym} \cite{Pretko:2016lgv,Pretko:2017fbf}
\be
\delta A_{\mu\nu}=\partial_\mu\partial_\nu\Phi\ ,
\label{fractonsym}\ee
where $\Phi(x)$ is a local scalar gauge parameter and the mass dimension of the tensor field is $[A_{\mu\nu}]=1$. Up to a field redefinition, the most general  local action compatible with power counting and invariant under \eqref{fractonsym} is
\be
\begin{split}
S_{inv}(a) =   \int d^4x &\left[
\partial_\mu A\partial^\mu A - \partial_\rho A_{\mu\nu}\partial^\rho A^{\mu\nu} +2A\partial_\mu\partial_\nu A^{\mu\nu} +2\partial^\lambda A_{\mu\lambda}\partial_\rho A^{\mu\rho}\right. \\
& \left. + a\left(\partial_\rho A_{\mu\nu}\partial^\rho A^{\mu\nu}-\partial^\lambda A_{\mu\lambda}\partial_\rho A^{\mu\rho}\right)\right]\ ,
\end{split}\label{Sinv}\ee
where $A(x)\equiv\eta^{\mu\nu}A_{\mu\nu}(x)$ and $\eta_{\mu\nu}=\mbox{diag}(-1,1,1,1)$ is the 4D Minkowski metric. 
The action \eqref{Sinv} depends on a dimensionless parameter, which we called ``$a$''. It is quite peculiar that a free quadratic theory shows a ``coupling'' constant which cannot be reabsorbed by a field redefinition. To our knowledge, the action \eqref{Sinv} is the only non interacting theory displaying this feature, with the exception of the abelian 3D Maxwell-Chern-Simons theory, where the coupling plays the role of a topological mass \cite{Deser:1981wh}. The choice of the $a$-parametrization of the action \eqref{Sinv} makes apparent the contact with LG \cite{Carroll.book}, which is reached for $a=0$~:
\be
S_{inv}(a=0)=S_{LG}\ ,
\label{}\ee
which obeys the stronger symmetry of infinitesimal diffeomorphisms 
\be
\delta_{diff}A_{\mu\nu}=\frac{1}{2}(\partial_\mu\xi_\nu+\partial_\nu\xi_\mu)\ .
\label{diff}\ee
The fracton symmetry \eqref{fractonsym} is recovered for the particular choice of the vector diff parameter
\be
\xi_\mu=(\partial_\mu\Phi)/2\ .
\label{}\ee
Notice that 
\be
\frac{\partial S_{inv}(a)}{\partial a} = \int d^4x\;
A_{\mu\nu}\left(\eta^{\mu\rho}\partial^2-\partial^\mu\partial^\rho\right)A_\rho^{\ \nu}\ ,
\label{aterm}\ee
and hence the ``$a$-term'' in the action $S_{inv}(a)$ \eqref{Sinv} is transverse. The gauge symmetry \eqref{fractonsym} need to be fixed in order to compute the propagators, and here we notice a peculiar behaviour. While for the stronger symmetry \eqref{diff} we promote the gauge parameter $\xi_\mu(x)$ to a ghost field  with antighost $\bar\xi_\mu(x)$ and Nakanishi-Lautrup Lagrange multiplier $b_\mu(x)$ \cite{Nakanishi:1966zz,Lautrup:1967zz} with
\be
\delta\bar\xi_\mu=b_\mu\ ,
\label{}\ee
in the weaker case \eqref{fractonsym} it is sufficient a scalar ghost $\phi(x)$ with antighost $\bar\Phi(x)$ and a multiplier $b(x)$ with
\be
\delta\bar\Phi=b\ .
\label{}\ee
Pursuing this second alternative enforces the following choice of dimensions
\be
[\Phi]=[\bar\Phi]=0\ \ ;\ \ [b]=1\ .
\label{}\ee

\subsection{Gauge fixing}

The most general term linear in $A_{\mu\nu}(x)$ and at most quadratic in $b(x)$ of maximal dimensionality is
\be
S_{gf}(k,k_1)=\int d^4x\left[
-b(\partial^\mu\partial^\nu A_{\mu\nu}+k_1\partial^2A)+\frac{k}{2}b\partial^2b
\right]\ ,
\label{Sgf}\ee
where $k$ and $k_1$ are two dimensionless gauge parameters. 
This term is peculiar, since if we try to eliminate the multiplier $b(x)$ by its equation of motion, we have
\be
\partial^\mu\partial^\nu A_{\mu\nu}+k_1\partial^2A-k\partial^2b=0\ ,
\label{gengaugecond}\ee
which substituted back in \eqref{Sgf} yields 
\be
S_{gf}=\int d^4x\left[
-\frac{1}{2}b(\partial^\mu\partial^\nu A_{\mu\nu}+k_1\partial^2A)\right]
=\frac{1}{2}S_{gf}(k=0,k_1)\ ,
\label{3.3}\ee
$i.e.$ it looks like we are forced into a Landau gauge $k=0$ choice. Following this suggestion we further consider the Landau gauge at $k_1=0$ and we have from \eqref{gengaugecond}
\be
\partial_\mu\partial_\nu A^{\mu\nu}=0\ ,
\label{}\ee
which closely resembles the Lorenz gauge condition of QED and furthermore it does not touch the transverse ``$a$-term'' \eqref{aterm} of the invariant action $S_{inv}(a)$ \eqref{Sinv}. For these reasons we will adopt the minimal Landau gauge \cite{Gambuti:2020onb}
\be
k=k_1=0\ ,
\label{minimalLandau}\ee
and the gauge fixed action in Fourier transform is
\bea
S &=& S_{inv}(a)+S_{gf} \nonumber\\
&=& \int d^4p   
\left[
p^2\At(p)\At(-p) 
- p^2\At_{\mu\nu}(p)\At^{\mu\nu}(-p) 
- 2\At(p)p_\mu p_\nu \At^{\mu\nu}(-p) \right. \label{Sgaugefixed} \\
&&\ \ \ \left.
+2p^\lambda p_\rho \At_{\mu\lambda}(p)\At^{\mu\rho}(-p)
+a \At_{\mu\nu}(p)(\eta^{\mu\rho}p^2-p^\mu p^\rho)\At_\rho^{\ \nu}(-p) 
+\bt(p) p^\mu p^\nu\At_{\mu\nu}(-p)\right]\ . \nonumber
\eea

\subsection{Propagators}\label{}

In order to compute the propagators of the theory, we rewrite the action $S_{inv}(a)$ \eqref{Sinv} as
\be
S_{inv}(a) = \frac{1}{2} \int d^4p\left[
\At_{\alpha\beta}(p)\tilde\Omega^{\alpha\beta,\mu\nu}(p)\At_{\mu\nu}(-p)\ ,
\right]
\label{Sinvp}\ee
where $\tilde\Omega^{\alpha\beta,\mu\nu}(p)$ satisfies
\be
\tilde\Omega^{\alpha\beta,\mu\nu}=\tilde\Omega^{\mu\nu,\alpha\beta}=
\tilde\Omega^{\beta\alpha,\mu\nu}=\tilde\Omega^{\alpha\beta,\nu\mu}\ ,
\label{}\ee
and can be expanded on five rank-4 tensors which we collectively denote \cite{Kugo:2014hja,Blasi:2015lrg,Amoretti:2013xya}
\begin{equation}
X_{\mu \nu, \alpha \beta} \equiv (A,B,C,D,E)_{\mu\nu,\alpha\beta}
\label{defX}\end{equation}
defined as
\begin{align}
    A_{\mu \nu, \alpha \beta} &= \frac{d_{\mu \nu} d_{\alpha \beta}}{3} \label{A}\\[10pt]
 B_{\mu \nu, \alpha \beta} &= e_{\mu \nu} e_{\alpha \beta} \label{B}\\[10pt]
  C_{\mu \nu, \alpha \beta} &= \frac{1}{2} \left(  d_{\mu \alpha} d_{\nu \beta} + d_{\mu \beta} d_{\nu \alpha} - \frac{2}{3} d_{\mu \nu} d_{\alpha \beta}  \right) \label{C}\\[10pt]
  D_{\mu \nu, \alpha \beta} &=  \frac{1}{2} \left(  d_{\mu \alpha} e_{\nu \beta} + d_{\mu \beta} e_{\nu \alpha} + e_{\mu \alpha} d_{\nu \beta} + e_{\mu \beta} d_{\nu \alpha}  \right)\label{D}\\[10pt]
  E_{\mu \nu, \alpha \beta} &= \frac{\eta_{\mu \nu} \eta_{\alpha \beta}}{4} \label{E} \; ,
\end{align}
where
\be
e_{\mu\nu}=\frac{p_\mu p_\nu}{p^2}\ \ ;\ \ d_{\mu\nu}=\eta_{\mu\nu} - e_{\mu\nu}
\label{edproj}\ee
are idempotent and orthogonal rank-2 projectors
\begin{equation}
    e_{\mu \lambda} {e^\lambda}_\mu = e_{\mu \nu}, \ \mathrm{}\   d_{\mu \lambda} {d^\lambda}_\nu = d_{\mu \nu}, \ \mathrm{}\     e_{\mu \lambda} {d ^\lambda}_\nu  =0 \: .  
\label{defed}\end{equation}
On the above basis, $\tilde\Omega^{\mu\nu,\alpha\beta}(p)$ in the action \eqref{Sinvp} writes
\be
\tilde\Omega^{\mu \nu , \alpha \beta} = t(p) A^{\mu \nu , \alpha \beta} + u(p) B^{\mu \nu , \alpha \beta} + v(p) C^{\mu \nu , \alpha \beta} + z(p) D^{\mu \nu , \alpha \beta} + w(p) E^{\mu \nu , \alpha \beta}\ ,
\label{Omegaexp}\ee
with
\be
t(p)=(2+a)p^2\ \ ;\ \ u(p)=0\ \ ;\ \ 
v(p)=(a-1)p^2\ \ ;\ \ z(p)=\frac{a}{2}p^2\ \ ;\ \ 
w(p)=0\ .
\label{Omegacoeff}\ee
The gauge fixed action $S$ \eqref{Sgaugefixed} can be written as
\bea
S &=& \int d^4p \left[
\At_{\mu\nu}\tilde\Omega^{\mu\nu,\alpha\beta}\At_{\alpha\beta} 
+ \bt(\At_{\mu\nu}\Lambda^{*\mu\nu}+\Lambda^{\mu\nu}\At_{\mu\nu})
\right] \nn\\
&=&\int d^4p\;
(\At_{\mu\nu}\ \ \bt)
\left(
\begin{array}{cc}
\tilde\Omega^{\mu\nu,\alpha\beta} & p^2e^{\mu\nu}/2 \\
p^2e^{\alpha\beta}/2&0
\end{array}
\right)
\left(
\begin{array}{c}
\At_{\alpha\beta} \\
\bt
\end{array}
\right)\ .
\label{Smatrix}\eea
The propagators are given by the inverse of the matrix appearing in \eqref{Smatrix}
\be
\left(
\begin{array}{cc}
\tilde\Omega^{\mu\nu,\alpha\beta} & p^2e^{\mu\nu}/2 \\
p^2e^{\alpha\beta}/2 & 0 
\end{array}
\right)
\left(
\begin{array}{cc}
\tilde{G}_{\alpha\beta,\rho\sigma} & \tilde{G}^*_{\alpha\beta} \\
\tilde{G}_{\rho\sigma} & \tilde{G} 
\end{array}
\right)
=
\left(
\begin{array}{cc}
\mathcal{I}^{\mu \nu}_{\ \ \rho \sigma}& 0 \\
0 & 1 
\end{array}
\right)\ ,
\label{propeq}\ee
where we parametrize the gauge propagator as
\be
\tilde{G}_{\alpha \beta,\rho\sigma} = \tilde{t}(p) A_{\alpha \beta,\rho\sigma}+ \tilde{u}(p) B_{\alpha \beta,\rho\sigma}+ \tilde{v}(p) C_{\alpha \beta,\rho\sigma} + \tilde{z}(p) D_{\alpha \beta,\rho\sigma} + \tilde{w}(p) E_{\alpha \beta,\rho\sigma}\ .
\label{Gtildeexp}\ee
We get (see Appendix A for the detailed calculations) 
\be
\tilde{t}(p)=\frac{1}{(2+a)p^2}\ \ ;\ \ 
\tilde{v}(p)=\frac{1}{(a-1)p^2}\ \ ;\ \ 
\tilde{z}(p)=\frac{2}{a p^2}\ . 
\label{coeffprop}\ee
As \eqref{coeffprop} clearly shows, the limit cases $a=0$, $|a|\rightarrow\infty$, $a=1$ and $a=-2$ should be considered separately. Let us recall that the parameter $a$ is the one which measures the deformation of the action $S_{inv}(a)$ \eqref{Sinv} with respect to LG.
\begin{enumerate}
\item $\bf a\rightarrow 0$
The action becomes that of LG and the symmetry is restored to the one shown in \eqref{diff}. Notice that the choice $a=0$ is protected by the enhanced symmetry \eqref{diff}. The propagator matrix is dominated by the $\tilde{z}(p)$ term in \eqref{Gtildeexp}~:
\be
\tilde{G}_{\alpha\beta,\mu\nu}(0)\propto \frac{2}{a p^2}D_{\alpha\beta,\mu\nu}\ .
\label{}\ee
\item $\bf |a|\rightarrow \infty$
The LG component of the action \eqref{Sinv} is suppressed and the propagator matrix becomes
\be
\tilde{G}_{\alpha\beta,\mu\nu}(\infty)\propto \frac{1}{a p^2}
(A_{\alpha\beta,\mu\nu}+C_{\alpha\beta,\mu\nu}+2D_{\alpha\beta,\mu\nu})\ .
\label{}\ee
Notice that
\be
\tilde{G}_{\alpha\beta,\mu\nu}(\infty)= \frac{1}{2a p^2}
(d_{\mu\alpha}d_{\nu\beta}+d_{\mu\beta}d_{\nu\alpha}) + 
\tilde{G}_{\alpha\beta,\mu\nu}(0)\ .
\label{}\ee
\item $\bf a\rightarrow 1$
Let us go back to the invariant action $S_{inv}(a)$ \eqref{Sinv} and see what happens when $a=1$. We find
\be
S_{inv}(a=1) = 
\int d^4p\; [p^2\At(p)\At(-p)-2\At(p)p^\mu p^\nu\At_{\mu\nu}(-p)
+p^\mu p^\nu \At_\mu^{\ \rho}(p)\At_{\rho\nu}(-p)]\ .
\label{Sinv1}\ee
Now, setting
\be
\tilde{\cal A}^\rho(p)\equiv p^\rho\At(p)- p^\nu\At_\nu^{\ \rho}(p)\ ,
\label{vectorfield}\ee
with
\be
\delta \tilde{\cal A}^\rho(p) = p^\rho p^2\tilde\Phi-p^\nu p_\nu p^\rho\tilde\Phi=0\ ,
\label{}\ee
we can rewrite \eqref{Sinv1} as
\be
S_{inv}(a=1)=\int d^4p\;[\tilde{\cal A}^\rho(p)\tilde{\cal A}_\rho(-p)]\ ,
\label{trivialaction}\ee
so that the action trivializes, since the field $\tilde{\cal A}^\rho(p)$ does not propagate, which is the defining property of the fracton quasiparticle. We shall comment on this in the concluding Section.
\item $\bf a\rightarrow -2$
Once again we reconsider the invariant action $S_{inv}(a)$ \eqref{Sinv} for this particular value of $a$, and setting
\be
A_{\mu\nu}(x)=\hat{A}_{\mu\nu}(x)+\frac{1}{4}\eta_{\mu\nu}A(x)\ ,
\label{}\ee
where $\hat{A}_{\mu\nu}(x)$ is the traceless component of $A_{\mu\nu}(x)$
\be
\hat{A}(x)=\hat{A}_\mu^{\ \mu}(x)=0\ ,
\label{}\ee
we find that $S_{inv}(a=-2)$ depends only on $\hat{A}_{\mu\nu}(x)$ and the trace $\hat{A}(x)$ disappears from the invariant action.
\end{enumerate}
Notice that, contrary to the $a\rightarrow 0$ case which is protected by the rising of a new symmetry \eqref{diff}, the cases $a=1$ and $a=-2$ are ``unstable'' under radiative corrections and therefore they correspond to a ``mild'' singularity which is there at the classical level but which does not survive the quantum fluctuations.

\section{Degrees of freedom (dof)}\label{}

Since the dof are gauge independent, the easiest way of computing them is in the Landau gauge through the equations of motion. From \eqref{Sgaugefixed} we have
\bea
\frac{\delta S}{\delta \bt} &=& p^\rho p^\sigma \At_{\rho\sigma}=0 \label{eomb}\\
\frac{\delta S}{\delta\At_{\mu\nu}} &=& 
2p^2\eta^{\mu\nu}\At-2p^2\At^{\mu\nu} -2\eta^{\mu\nu}p^\rho p^\sigma\At_{\rho\sigma}
-2p^\mu p^\nu\At + 2(p^\mu p^\tau\At^\nu_{\ \tau}+p^\nu p^\tau\At^\mu_{\ \tau}) \nonumber\\
&& +a\;(2p^2\At^{\mu\nu}- p^\mu p^\tau\At^\nu_{\ \tau}-p^\nu p^\tau\At^\mu_{\ \tau})
+p^\mu p^\nu \bt =0\ .\label{eomA}
\eea
From \eqref{eomA} we find
\bea
\eta_{\mu\nu}\frac{\delta S}{\delta\At_{\mu\nu}} &=& 
(4+2a)(p^2\At-p^\rho p^\sigma\At_{\rho\sigma})+p^2\bt =0 \\
e_{\mu\nu}\frac{\delta S}{\delta\At_{\mu\nu}} &=& p^2\bt =0 \\
p_\nu \frac{\delta S}{\delta\At_{\mu\nu}} &=& 
a\;(p^2p_\nu\At^{\mu\nu}-p^\mu p^\rho p^\sigma\At_{\rho\sigma})+p^\mu p^2\bt=0\ ,
\eea
which, taking into account also the gauge condition \eqref{eomb}, imply
\bea
p^2\bt &=&0 \label{5.6}\\
p^\rho p^\sigma\At_{\rho\sigma} &=& 0 \label{5.7}\\ 
a p^2p_\nu\At^{\mu\nu}&=&0 \label{5.8}\ .
\eea
A remark is in order concerning the case $a=0$, which corresponds of having only gravitons in the theory, as we see from $S_{inv}(a)$ \eqref{Sinv}. For $a=0$ we have only the constraint \eqref{5.7} on the symmetric tensor field $A_{\mu\nu}(x)$, which therefore would display 9 dof. Now, we know that the dof of massless LG are, instead, 6 \cite{Hinterbichler:2011tt,deRham:2014zqa}. It is interesting and instructive to go back to the origin of this mismatch. We already noticed that LG is a particular solution of the fracton symmetry \eqref{fractonsym}, which is protected by the stronger infinitesimal diffeomorphism symmetry \eqref{diff}. In other words, for $a\rightarrow 0$ the symmetry switches from the weaker \eqref{fractonsym} to the stronger \eqref{diff}. This transition is not continuous as far as the gauge parameter is concerned, because it changes from being a scalar (for the fracton symmetry \eqref{fractonsym}) to become a vector field (for the infinitesimal diffeomorphism \eqref{diff}). As a consequence, the ``scalar'' gauge fixing realised in $S_{gf}$ \eqref{Sgf} is not a proper gauge fixing for the $a=0$ case, and one should choose instead a ``vectorial'' gauge fixing, as done in \cite{Gambuti:2021meo,Blasi:2015lrg} for LG. This is the physical explanation of the divergence of the propagator \eqref{coeffprop} for $a\rightarrow 0$.
In the case $a\neq 0$, let us define
\be
\tilde{J}_\rho\equiv p^\sigma\At_{\rho\sigma}\ ,
\label{}\ee
which by \eqref{5.7} is a conserved current, $i.e.$
\be
p^\rho\tilde{J}_\rho=0\ ,
\label{}\ee
which implies that the current $\tilde{J}_\rho$ is of the form
\be
\tilde{J}_\rho=\epsilon_{\rho\mu\nu\lambda}p^\mu\tilde{B}^{\nu\lambda}
\label{tildeJexpr}\ee
for a generic antisymmetric tensor $\tilde{B}^{\nu\lambda}$.
Substituting back $\tilde{J}_\rho$  in \eqref{5.8} yields
\be
p^2\tilde{J}_\mu = p^2 \epsilon_{\mu\rho\nu\lambda}p^\rho\tilde{B}^{\nu\lambda}=0\ ,
\label{}\ee
which also implies
\be
p^2\tilde{B}_{\rho\lambda}=p_\rho\tilde{B}_\lambda - p_\lambda\tilde{B}_\rho\ ,
\label{}\ee
or
\be
\tilde{B}_{\rho\lambda}=\frac{1}{p^2}(p_\rho\tilde{B}_\lambda - p_\lambda\tilde{B}_\rho)\ ,
\label{}\ee
and hence, from \eqref{tildeJexpr}
\be
\tilde{J}_\mu=0\ ,
\label{}\ee
or
\be
p^\sigma\At_{\rho\sigma}=0\ ,
\label{}\ee
which represents four conditions. Therefore at $a\neq 0$ we find 6 dof, except the case $a=-2$ where the tensor becomes traceless and the dof reduce to 5.


\section{Massive theory}\label{}

To complete the analysis, we consider the massive case, and the generic mass term that can be added to the invariant action is \cite{Gambuti:2021meo,Blasi:2015lrg,Blasi:2017pkk}
\bea
S_m &=& \int d^4p\; \left(\frac{m_1}{4}\At^{\mu\nu}\At_{\mu\nu} + \frac{m_2}{8}\At^2\right)          \nonumber\\
&=& \int d^4p\; \frac{1}{2} \At_{\mu\nu}
(m_1  \mathcal{I}^{\mu \nu, \rho \sigma} + m_2 E^{\mu\nu,\rho\sigma})
\At_{\rho\sigma}\ ,
\label{}\eea
where  $\mathcal{I}^{\mu \nu, \rho \sigma}$ and $E^{\mu\nu,\rho\sigma}$ are the rank-4 tensors defined in \eqref{identity} and \eqref{E}. Consequently, the tensor 
$\tilde\Omega^{\mu\nu,\rho\sigma}$ \eqref{Sinvp} is modified as
\be
\tilde\Omega^{\mu\nu,\rho\sigma}(m_1,m_2)=\tilde\Omega^{\mu\nu,\rho\sigma} + m_1  \mathcal{I}^{\mu \nu, \rho \sigma} + m_2 E^{\mu\nu,\rho\sigma}\ ,
\label{}\ee
and, by Eqs. \eqref{identity} and \eqref{Omegaexp}, we can write
\be
\tilde\Omega^{\mu\nu,\rho\sigma}(m_1,m_2)= (t(p)+m_1)A^{\mu\nu,\rho\sigma} + m_1B^{\mu\nu,\rho\sigma} +(v(p) + m_1)C^{\mu\nu,\rho\sigma} + (z(p)+m_1) D^{\mu\nu,\rho\sigma} +m_2E^{\mu\nu,\rho\sigma}\ .
\label{}\ee
Since our aim here is to investigate the propagator structure of the invariant theory, we shall keep the modification with respect to the massless action to a minimum and accordingly we shall maintain the minimal Landau gauge fixing, given by $S_{gf}(k=k_1=0)$ \eqref{Sgf}.
To compute the propagators we have to find a matrix
\be
\left(
\begin{array}{cc}
\tilde{G}_{\alpha\beta,\rho\sigma}(m_1,m_2) & \tilde{G}^*_{\alpha\beta}(m_1,m_2) \\
\tilde{G}_{\rho\sigma}(m_1,m_2) & \tilde{G}(m_1,m_2)
\end{array}
\right)
\label{}\ee
which obeys
\be
\left(
\begin{array}{cc}
\tilde\Omega^{\mu\nu,\alpha\beta}(m_1,m_2) & p^2e^{\mu\nu}/2 \\
p^2e^{\alpha\beta}/2 & 0 
\end{array}
\right)
\left(
\begin{array}{cc}
\tilde{G}_{\alpha\beta,\rho\sigma}(m_1,m_2) & \tilde{G}^*_{\alpha\beta}(m_1,m_2) \\
\tilde{G}_{\rho\sigma}(m_1,m_2) & \tilde{G}(m_1,m_2)
\end{array}
\right)
=
\left(
\begin{array}{cc}
\mathcal{I}^{\mu \nu}_{\ \ \rho \sigma}& 0 \\
0 & 1 
\end{array}
\right)\ .
\label{massivematrixpropeq}\ee
Parametrising $\tilde{G}_{\alpha\beta,\rho\sigma}(m_1,m_2)$ as
\be
\tilde{G}_{\alpha\beta,\rho\sigma}(m_1,m_2)=
\tilde{T}A_{\mu\nu,\rho\sigma} + \tilde{U}B_{\mu\nu,\rho\sigma} +
\tilde{V}C_{\mu\nu,\rho\sigma} + \tilde{Z}D_{\mu\nu,\rho\sigma} +
\tilde{W}E_{\mu\nu,\rho\sigma}
\label{massiveprop}\ee
we find (the details of the calculation are in the Appendix)
\be
m_2=0\ , 
\label{}\ee
\be
\tilde{G}(m_1,m_2)= -\frac{2m_1}{p^4}\ 
\label{}\ee
and
\bea
\tilde{T}(p;a,m_1) &=& \frac{1}{(2+a)p^2+m_1} \label{6.18} \\
\tilde{V}(p;a,m_1) &=& \frac{1}{(a-1)p^2+m_1} \label{6.19}\\
\tilde{Z}(p;a,m_1) &=& \frac{1}{ap^2/2+m_1}\ .\label{6.20}
\eea
A few considerations are in order. First of all, we observe that in the Landau gauge a separate mass term for the trace of $A_{\mu\nu}(x)$ is forbidden, in close analogy to what happens in massive LG \cite{Gambuti:2021meo}. Moreover, having introduced a mass term, we have the new propagator \eqref{massiveprop} with coefficients \eqref{6.18}-\eqref{6.20}, which we would like not to have tachyonic poles \cite{Blasi:2015lrg}. With our choice of the metric $\eta_{\mu\nu}=\mbox{diag}(-1,1,1,1)$ we need that the coefficient of $p^2$ and the mass term have the same sign, and this identifies  two allowed regions in the $(a,m_1)$-plane.
\begin{figure}[!h]
\centering
\includegraphics[scale=.2]{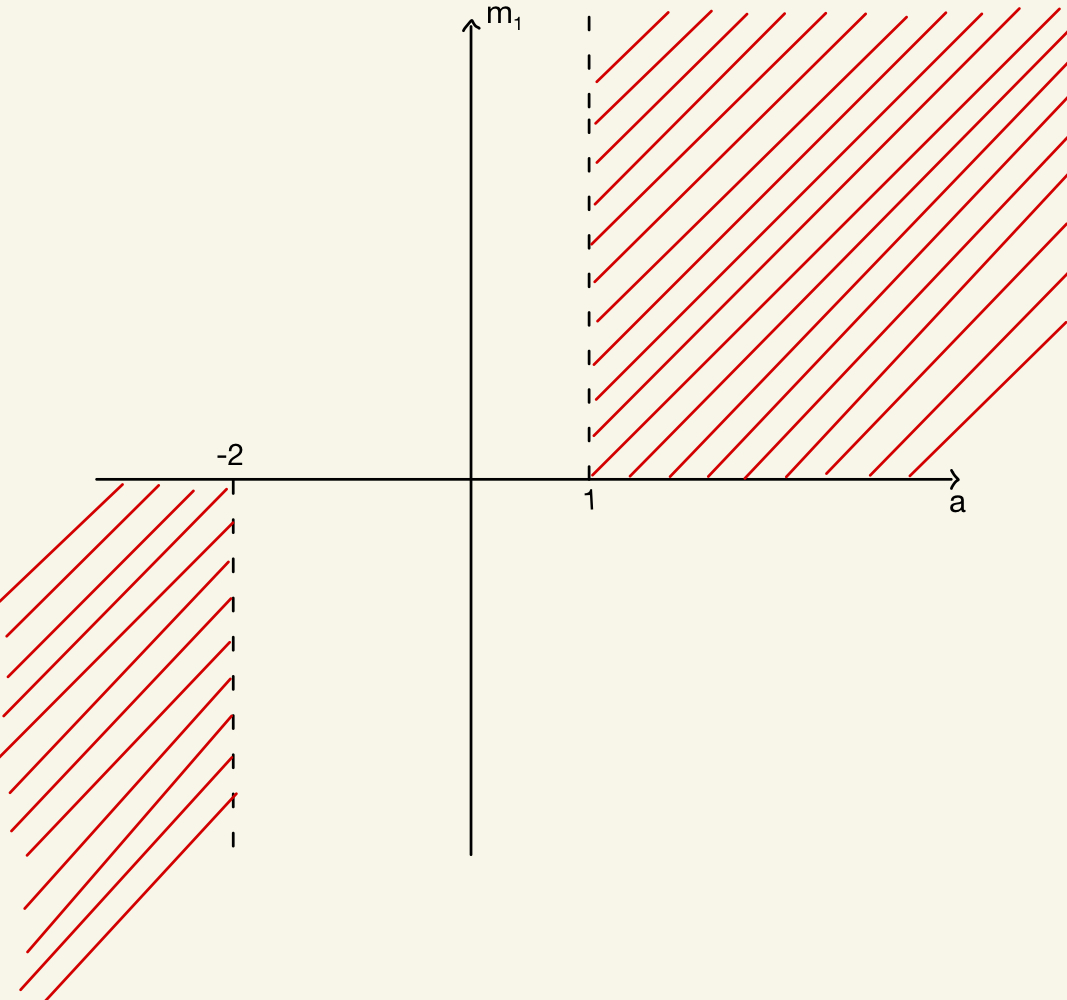}
\caption{Allowed regions in the $(a,m_1)$-plane}
\end{figure}
Therefore the $a$-segment $(-2,1)$ is not physically reachable and consequently in the massive case we cannot reach $a=0$, $i.e.$ the point where fractons disappear and only gravitons are present. This signals that, as already pointed out in the study of the dof of the theory, the case $a=0$ is somewhat singular for the gauge choice \eqref{gengaugecond}, due to the fact that the gauge symmetry at $a=0$ displays a vector, rather than scalar, parameter. The problem of massive LG has a longstanding literature, and a unified theoretical approach to both massive fractons and massive gravitons is still lacking.
The number of dof in the massive case can be computed in complete analogy to the massless one, and we find two constraints
\bea
p_\rho p_\sigma \At^{\rho\sigma} &=& 0 \\
\left(ap^2+\frac{m_1}{2}\right)p_\mu\At^{\mu\nu} &=&0\ ,\label{6.22}
\eea
but now $a\neq 0$ and therefore we get from \eqref{6.22} four constraints and hence six dof, like in the massless case.
\\

We finally remark that in the Landau gauge we have a functional equation which controls the mass term which breaks the gauge symmetry \eqref{fractonsym}, indeed we find
\be
\delta  S(m)=\frac{m_1}{2}\int d^4x\; A^{\mu\nu}\partial_\mu\partial_\nu\Phi
=\frac{m_1}{2}\int d^4x\; \Phi\partial_\mu\partial_\nu A^{\mu\nu}
=-\frac{m_1}{2}\int d^4x\; \Phi \frac{\delta}{\delta b} S(m)\ ,
\label{}\ee
hence
\be
\left(
\delta +\frac{m_1}{2}\int d^4x\; \Phi \frac{\delta}{\delta b}\right) S(m)\ .
\label{}\ee


Notice that if in \eqref{6.18}-\eqref{6.20} we parametrize $m_1$ by means of the $a$-parameter as follows
\be
m_1=\frac{a}{|a|}m^2\ ,
\label{}\ee
to avoid tachyons with imaginary mass when
\begin{itemize}
\item {\bf $\mathbf{a>0}$}: from $\tilde{V}$ \eqref{6.19} we find $a\geq1$\ ,
\item {\bf $\mathbf{a<0}$}: from $\tilde{T}$ \eqref{6.18} we find $a\leq -2$\ ,
\end{itemize}
therefore the region $a\in(-2,1)$ is not physically reachable, as in Figure 1.

\section{Summary and discussion}\label{}

In this paper we considered the theory of a symmetric tensor field $A_{\mu\nu}(x)$ in 4D, characterized by a particular infinitesimal diffeomorphism transformation, namely that whose vectorial parameter is a 4-divergence of a scalar. This weaker transformation identifies as the most general invariant action the local functional $S_{inv}$ \eqref{Sinv}, which depends on a dimensionless parameter, which we called ``$a$'', and which cannot be reabsorbed by a redefinition of $A_{\mu\nu}$. This is unusual for a quadratic theory and, to our knowledge, unique in 4D, the most known example of a quadratic theory depending on a true parameter being Maxwell Chern Simons theory in 3D, where the parameter plays the role of topological mass for the gauge vector field. This opens the question of the interpretation of the $a$-parameter, besides the obvious fact of tuning the invariant action which coincides with LG when $a=0$ to suppress gravitons. The fact of dealing with a scalar gauge parameter suggests a gauge fixing governed by a  Nakanishi-Lautrup scalar multiplier implementing a gauge condition which, as customary for symmetric rank-2 tensor fields \cite{Gambuti:2020onb,Blasi:2015lrg}, depends on two gauge parameters. We show how the Landau gauge choice is the natural one and that the value $a=0$, which would correspond to pure LG, is forbidden for two reasons. A technical one (the propagator diverge at $a\rightarrow 0$) and a physical one (wrong counting of the degrees of freedom, which turn out to be nine instead of six, as expected in LG). We interpreted this mismatch with the fact that the symmetry of pure LG is governed by a vector gauge parameter \eqref{diff} rather than a scalar one as in \eqref{fractonsym}. As a consequence, the gauge fixing needs a vector Nakanishi-Lautrup multiplier as done in \cite{Gambuti:2021meo,Gambuti:2020onb,Blasi:2015lrg,Blasi:2017pkk,Blasi:2008gt} and not a scalar one as in \eqref{Sgf}. A final, important comment concerns the relation between the theory studied in this paper and the fracton excitations. The first link is of course given by the symmetry \eqref{fractonsym}, which almost defines the fracton theory \cite{Pretko:2017fbf}, but does not uniquely fixes the fractons as the only quasiparticles of the theory, but rather a mixed phase of fractons and gravitons. What is really peculiar of fractons is the absence of mobility \cite{Chamon:2004lew,Vijay:2015mka,Vijay:2016phm}. Now, if fractons are localized, we may propose a different (and complementary) interpretation of the $a$-parameter. In Section (2.3) we noticed that the case $a=1$ we can write the invariant action in terms of the vector field \eqref{vectorfield}
\be
{\cal A}_\rho(x)\equiv \partial_\rho A(x)- \partial^\nu A_{\nu\rho}(x)
\label{calA}\ee
 as \eqref{trivialaction}
\be
S_{inv}(a=1)=\int d^4x\; {\cal A}_\rho {\cal A}^\rho\ ,
\label{7.1}\ee
which does not propagate and is also invariant since
\be
\delta {\cal A}_\rho=\partial_\rho(\Box\Phi)-\partial^\nu (\partial_\nu\partial_\rho\Phi) \textcolor{red}{=0}\ .
\label{}\ee
The action $S_{inv}(a)$ in (2.2) describes a class of theories parametrized by $a$ and invariant under the ``fracton'' symmetry (2.1). 
\begin{figure}[H]
\centering
\includegraphics[scale=.3]{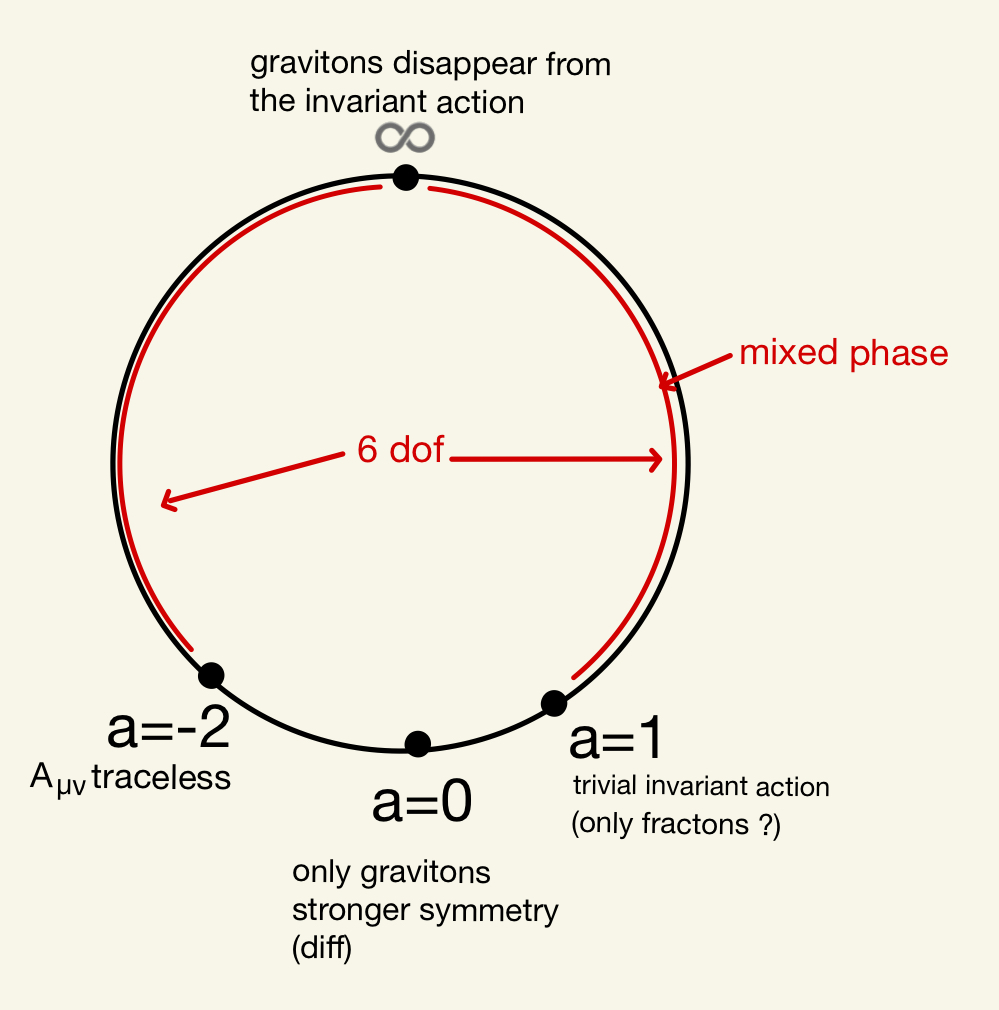}
\caption{Allowed values of the $a$-parameter}
\end{figure}
Amongst them, we observe that the one characterized by $a=1$ has a peculiar property: it is possible to make the linear redefinition \eqref{calA} which trivializes the action $S_{inv}(a=1)$ to \eqref{7.1} for a non propagating vector field. Now, interpreting the fracton property of absence of mobility as a non propagating excitation, we guess that $S_{inv}(a=1)$ might be the purely fracton action. For $a\rightarrow\infty$ the graviton disappears in the sense that the LG component of the action is suppressed in favour of the dominating $a$-component which, alone, is an invariant action for a propagating symmetric tensor field and, hence, cannot be identified as a fracton quasiparticle, which, instead, should not propagate. The case $a=1$, which is our candidate for a pure fracton theory, corresponds to an extension of the LG action, not to its suppression. The allowed values of the $a$-parameter are summarized in Figure 2.
Now define the parameter $\mu$ as
\be
\mu\equiv\left| \frac{a-1}{a}\right|\ ,
\label{mu}\ee
which measures the ``rate of propagation'' of the ``particle'' described by the action $S_{inv}(a)$ \eqref{Sinv}. In fact, for $a=1$ we have $\mu=0$, and the mode is localized, while for $a\rightarrow0$ we have $\mu\rightarrow\infty$, $i.e.$ the gravitons propagate freely. Finally for $a=-2$, we have $\mu=\frac{3}{2}$, and for $|a|\rightarrow\infty$ we have $\mu\rightarrow 1$. In terms of $\mu$ we have the situation displayed in Figure 3.
\begin{figure}[H]
\centering
\includegraphics[scale=.1]{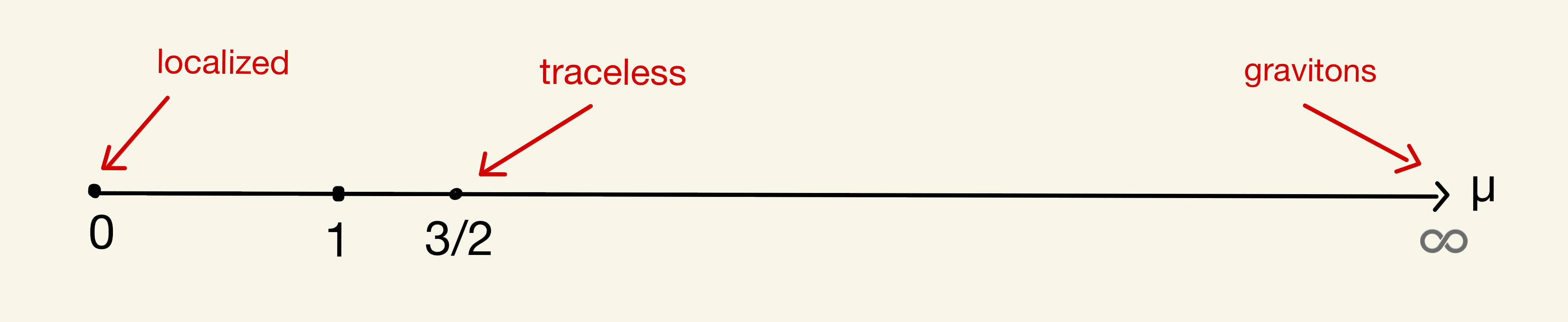}
\caption{Rate of propagation $\mu$}
\label{mobilityfig}\end{figure}
We have the quite uncommon situation where the action $S_{inv}(a)$ \eqref{Sinv}, which is only quadratic, contains a dimensionless parameter (``$a$''). In Section (2.3) we saw that the resulting propagators become $a$-dependent and we isolated three points, namely $a=0$, $a=1$ and $a=-2$, where these values simulate an infrared divergence which is the signal of a ``phase transition''. Under this respect, the only clear interpretation is for $a=0$, where the invariant action coincides with that of LG and there is a shift in the symmetry from the weaker scalar gauge \eqref{fractonsym} to the full infinitesimal diffeomorphism \eqref{diff}. In this sense we may say that at $a=0$ we have only gravitons.
Now the question is: if we eliminate the gravitons, $i.e.$ $a\rightarrow\infty$  what is left ? In this limit we have an independent term which is invariant under the weaker symmetry, but this term really identifies only an equivalence class and it  is within this class that we have to make a choice for the "fracton" action.
What we propose is strictly related to a peculiar property of fractons, $i.e.$ the fact that they are localized and do not propagate. Hence it seems natural to assume that at $a=1$, where the action trivializes and is written in terms of the non propagating invariant vector field ${\cal A}_\rho(x)$ \eqref{calA}, we have only fractons.
The definition of the rate of propagation \eqref{mu} allows us to 
make contact with the ``phase transition'' picture: suppose $a=\frac{T_c-T}{T_c}$ for some critical temperature $T_c$, then 
$\mu=\left|\frac{T}{T-T_c}\right|$ and we have $\mu=0$ for $T=0 K$ while $\mu\rightarrow\infty$ when $T\rightarrow T_c$. The case $a=-2$ means $T=3T_c$. If this picture is correct, we should have a critical temperature $T_c\neq 0K$ where we have a phase of only gravitons.
The singular point $a=-2$ has an immediate interpretation since the trace of the tensor field disappears from the invariant action and we loose one degree of freedom. To summarize, at $a=0$ we have only gravitons and at $a=1$ only fractons.
The hypothesis about the presence of a critical temperature T where the phase transition from fractons to gravitons takes place is simply a ``hypothesis'' which fits in our scheme but for which we have no explicit evidence
We conclude with a remark concerning the choice we made for the gauge fixing. The reason for choosing a scalar gauge condition instead of a vector one, as it happens in LG, which also is described by a symmetric tensor field, is due to the fact that the gauge parameter of the transformation \eqref{fractonsym} is a scalar function. Hence only one gauge degree of freedom needs to be fixed by means of a scalar constraint and, for a symmetric tensor field $A_{\mu\nu}$, the most general one is that implemented by \eqref{3.3}:
\begin{equation}
\partial^\mu\partial^\nu A_{\mu\nu}+k_1\partial^2A=0,
\label{scalargaugecond}\end{equation}
which generalizes the Lorenz gauge condition for the vector gauge field $A_\mu$
\begin{equation}
\partial^\mu A_\mu=0.
\end{equation}
For $a=0$, the action reduces to that of LG, which is not uniquely determined by the ``fracton'' transformation \eqref{fractonsym} but by the infinitesimal diffeomorphism transformation \eqref{diff}, which is a gauge transformation with a vector parameter, which hence necessarily needs a vector gauge condition, whose most general form is
\begin{equation}
\partial^\nu A_{\mu\nu}+k_1\partial_\mu A=0.
\label{vectorgaugecond}\end{equation}
The question might arise: what if we chose from the beginning a vector gauge condition, for a generic parameter $a$~? Our guess is that it would be correct for $a=0$ (which is LG), but it would overcount  the gauge degrees of freedom for $a\neq 0$, and we expect that this would reflect in a different (and presumably wrong) counting of the theory degrees of freedom. In any case, it should not be the first choice, when dealing with a gauge transformation like \eqref{fractonsym} characterized by a scalar parameter. 
As a first step towards a comparison of the two approaches, in Appendix A.1 we compute the propagators also for the vector gauge condition \eqref{vectorgaugecond}, but this field theoretical issue would lead beyond the scope of this Letter, which is that of studying the degrees of freedom of a theory of a symmetric tensor field parametrized by $a$, and finding a criterion for picking up amongst the whole class of theories, the one which might describe the fractons, whose defining property is that of being quasiparticles with vanishing or limited mobility. Our candidate for the fracton theory is $S_{inv}(a=1)$ \eqref{Sinv}, which can be reduced to the trivial action \eqref{7.1} for the non propagating vector field \eqref{calA}.
This, at the moment, seems to be the main feature to identify the fractons. Following this hint, from \eqref{mu} we see that for $a\rightarrow\infty$ the mobility $\mu\rightarrow 1$, which is a small value compared to the gravitons mobility ($a\rightarrow 0 \Rightarrow\mu\rightarrow\infty$). If we wish to give a name to the quasiparticles with $a\neq 1$ (and $a\neq 0$), we might call them ``quasifractons''. We would like to point out that our picture is only a possibility, and that further investigation is needed to arrive at a well established physical identification of the whole class of quasiparticles described by the action \eqref{Sinv}.

\section*{Acknowledgments}

We thank Andrea Amoretti, Erica Bertolini, Daniele Musso and Giandomenico Palumbo for enlightening discussions. This work has been partially supported by the INFN Scientific Initiative GSS: ``Gauge Theory, Strings and Supergravity''. 

\appendix

\section{Propagators}

\subsection{Massless case}

The $X_{\mu \nu, \alpha \beta}$ tensors \eqref{defX} have the symmetries
\begin{equation}
X_{\mu \nu, \alpha \beta}  = X_{\nu \mu, \alpha \beta}     = X_{\mu \nu, \beta \alpha }    = X_{\alpha \beta, \mu \nu}      \; ,
\end{equation}
and the following properties hold:
\begin{itemize}
    \item 
    decomposition of the rank-4 tensor identity 
     \begin{equation}
            \mathcal{I}_{\mu \nu, \rho \sigma} = \frac{1}{2} (\eta_{\mu \rho} \eta_{\nu \sigma} + \eta_{\mu \sigma} \eta_{\nu \rho}) 
\label{identity} \end{equation}
    \begin{equation}
    A_{\mu \nu , \alpha \beta} + B_{\mu \nu , \alpha \beta} + C_{\mu \nu , \alpha \beta} + D_{\mu \nu , \alpha \beta} =  \mathcal{I}_{\mu \nu , \alpha \beta}\ ;
    \label{idempotency}\end{equation}
    \item idempotency~:
    \begin{equation}
    X_{\mu\nu}^{\ \ \rho\sigma}X_{\rho\sigma,\alpha\beta}=X_{\mu\nu,\alpha\beta}\ ;
    \label{idempotency}\end{equation}
    \item orthogonality of $A$, $B$, $C$ and $D$~:
    \begin{equation}
    X_{\mu \nu , \alpha \beta} {{X^\prime}^{\alpha \beta}}_{\rho \sigma} = 0\ \ \mbox{if}\
    (X,X^\prime)\neq E\ \mbox{and}\ X\neq X^\prime\ ;
    \label{orthogonality}\end{equation}
    \item contractions with $E$~:
    \begin{align} 
    A_{\mu \nu , \alpha \beta} {E^{\alpha \beta}}_{\rho \sigma} 
    &= \frac{d^{\mu \nu} \eta_{\rho \sigma}}{4}\label{AE}\\[10pt] 
    B_{\mu \nu , \alpha \beta} {E^{\alpha \beta}}_{\rho \sigma} 
    &= \frac{e^{\mu \nu} \eta_{\rho \sigma}}{4}\label{BE}\\[10pt]
    C_{\mu \nu , \alpha \beta} {E^{\alpha \beta}}_{\rho \sigma}
    &=D_{\mu \nu , \alpha \beta} {E^{\alpha \beta}}_{\rho \sigma}=0\label{CE-DE}\ .    \end{align}
\end{itemize}
From \eqref{propeq} we have
\bea
\tilde\Omega^{\mu\nu,\alpha\beta}\tilde{G}_{\alpha\beta,\rho\sigma}
+\frac{1}{2}p^2e^{\mu\nu}\tilde{G}_{\rho\sigma}
&=& \mathcal{I}^{\mu \nu}_{\ \ \rho \sigma} \label{4.24} \\
\tilde\Omega^{\mu\nu,\alpha\beta}\tilde{G}^*_{\alpha\beta}     
+\frac{1}{2}p^2e^{\mu\nu}\tilde{G}
&=& 0 \label{4.25} \\
p^2e^{\alpha\beta}\tilde{G}_{\alpha\beta,\rho\sigma}
&=& 0 \label{4.26}\\
\frac{1}{2}p^2e^{\alpha\beta}\tilde{G}^*_{\alpha\beta} 
&=&1\ . \label{4.27}
\eea
From \eqref{4.27} we immediately get
\be
\tilde{G}^*_{\alpha\beta} =\tilde{G}_{\alpha\beta}= \frac{2}{p^2}e_{\alpha\beta}\ ,
\label{}
\ee
which, substituted in \eqref{4.25} and taking into account \eqref{Omegaexp} and \eqref{Omegacoeff}, gives
\be
\tilde{G}=0\ .
\label{}\ee
Using the properties of the rank-2 projectors $e_{\mu\nu}$ and $d_{\mu\nu}$ \eqref{edproj} and of the rank-4 tensors $X_{\mu\nu,\alpha\beta}$ \eqref{defX}, from \eqref{4.26}  we have
\be
\tilde{u}(p)=\tilde{w}(p)=0\ .
\label{}\ee
The other coefficients of the gauge propagator $\tilde{G}_{\alpha \beta,\rho\sigma}$ \eqref{Gtildeexp} are determined by \eqref{4.24}, which becomes
\be
t\tilde{t}A^{\mu\nu}_{\ \ \rho\sigma}+v\tilde{v}C^{\mu\nu}_{\ \ \rho\sigma}+
z\tilde{z}D^{\mu\nu}_{\ \ \rho\sigma}+e^{\mu\nu}e_{\rho\sigma}=\mathcal{I}^{\mu \nu}_{\ \ \rho \sigma}=A^{\mu\nu}_{\ \ \rho\sigma}+B^{\mu\nu}_{\ \ \rho\sigma}+C^{\mu\nu}_{\ \ \rho\sigma}+D^{\mu\nu}_{\ \ \rho\sigma}\ ,
\label{}\ee
hence
\be
t\tilde{t}=v\tilde{v}=z\tilde{z}=1\ ,
\label{}\ee
$i.e.$
\be
\tilde{t}(p)=\frac{1}{(2+a)p^2}\ \ ;\ \ 
\tilde{v}(p)=\frac{1}{(a-1)p^2}\ \ ;\ \ 
\tilde{z}(p)=\frac{2}{a p^2}\ . 
\label{}\ee
The symmetry \eqref{fractonsym}, besides many other physical peculiarities, has also the additional one of giving a double possibility of gauge fixing it, by means of the scalar gauge condition \eqref{scalargaugecond} or the vector one \eqref{vectorgaugecond}. In this paper we adopted the scalar choice by counting the gauge degrees of freedom: the symmetry \eqref{fractonsym} involves a scalar gauge parameter. Hence, the scalar gauge choice \eqref{scalargaugecond} is the appropriate one. Nevertheless, in order to be able to write the propagators of the theory, one might adopt the vector gauge condition \eqref{vectorgaugecond} as well, which is the one relevant in LG and which corresponds to the stronger infinitesimal diff transformation \eqref{diff}, which is not a symmetry of the theory studied in this paper (the action \eqref{Sinv} {\bf is not} invariant under the gauge transformation \eqref{diff}). Hence, the vectorial gauge condition \eqref{vectorgaugecond} is not the correct one. This opens an interesting, purely field theoretical, question, on the nature of the gauge fixing procedure. Going back to Faddeev and Popov, gauge fixing a theory is a much more subtle task than adding a term to the action in order to being able to invert the quadratic part of the action and write down the propagators. Gauge fixing a gauge field theory means intersecting the gauge orbits and choosing, for each orbit, one representative, and possibly only one. We know that this is not possible, due to the Gribov problem, but what is sure is that a scalar gauge parameter needs one and only one gauge condition. Choosing, as it would be the case for the vector condition \eqref{vectorgaugecond}, four conditions instead of one is more than enough for writing the propagators, but it does not correspond to intersecting the gauge orbits defined by \eqref{fractonsym} only once. Hence, the existence of propagators is granted, as we shall show in a moment, but the counting of the degrees of freedom would be presumably wrong. Nevertheless, for the sake of completeness, we now compute  the propagators corresponding to the vector gauge fixing \eqref{vectorgaugecond}. 
The gauge fixing term in this case is
\begin{equation}\label{eq:gaugefixing}
S_{gf} = \int d^4x \left\{ b^\mu\left[\partial^\nu A^{\mu\nu} +k_1\partial_\mu A\right] + \frac{k}{2}b^\mu b_\mu\right\}\ ,
\end{equation}
where $b^\mu$ is a vector Lagrange multiplier implementing the condition \eqref{vectorgaugecond} and $(k,k_1)$ are two gauge parameters. 
In Fourier transform the total action reads
\be
S =
\int d^4p\;
(\At_{\mu\nu}\ \ \bt_\mu)
\left(
\begin{array}{cc}
\tilde\Omega^{\mu\nu,\alpha\beta} & \tilde\Lambda^{\star\mu\nu,\alpha} \\
\tilde\Lambda^{\mu,\alpha\beta}&\tilde{H}^{\mu\alpha}
\end{array}
\right)
\left(
\begin{array}{c}
\At_{\alpha\beta} \\
\bt_\alpha
\end{array}
\right)\ ,
\label{}\ee
where $\tilde\Omega^{\mu\nu,\alpha\beta}$ is still given by \eqref{Omegaexp}
and
\bea
\tilde\Lambda^{\mu,\alpha\beta} &=&  
-\frac{i}{4}(d^{\alpha\mu}p^\beta+d^{\beta\mu}p^\alpha)
-\frac{i}{2}k_1d^{\alpha\beta}p^\mu-\frac{i}{2}(1+k_1)e^{\alpha\beta}p^\mu
\\
\tilde{H}_{\mu\nu} &=& 
\frac{k}{2}(d_{\mu\nu}+e_{\mu\nu})\ .
\eea
The matrix of propagators should satisfy
\be
\left(
\begin{array}{cc}
\tilde\Omega^{\mu\nu,\alpha\beta} & \tilde\Lambda^{\star\mu\nu,\alpha} \\
\tilde\Lambda^{\mu,\alpha\beta}&\tilde{H}^{\mu\alpha}
\end{array}
\right)
\left(
\begin{array}{cc}
\tilde{G}_{\alpha\beta,\rho\sigma} & \tilde{G}_{\alpha\beta,\rho} \\
\tilde{G}^*_{\rho\sigma,\alpha} & \tilde{G}_{\alpha\rho}
\end{array}
\right)
=
\left(
\begin{array}{cc}
\mathcal{I}^{\mu \nu}_{\ \ \rho \sigma}& 0 \\
0 & \eta_{\mu\rho}
\end{array}
\right)\ ,
\label{propeq}\ee
where the gauge propagator is parametrized as \eqref{Gtildeexp}, and
\bea
\tilde G_{\mu\nu,\alpha} &=& 
i[\tilde{f}(p)(d_{\mu\alpha}p_\nu+d_{\nu\alpha}p_\mu)+\tilde{g}(p)d_{\mu\nu}p_\alpha+\tilde{l}(p)e_{\mu\nu}p_\alpha]\\
\tilde G_{\mu\nu} &=& \tilde{r}(p)d_{\mu\nu}+\tilde{s}(p)e_{\mu\nu}\ .
\eea
After long but straightforward calculations we get
\bea
\tilde{t}(p) &=& \frac{1+4k_1}{(1+k_1)(a+2)p^2}\ ;\
\tilde{u}(p) = \frac{k_1(1+4k_1)-2k(a+2)}{(1+k_1)^2(a+2)p^2}\ ;\ 
\tilde{v}(p) = \frac{1}{(a-1)p^2} \\
\tilde{z}(p) &=& \frac{4k}{(2ka-1)p^2}\ ;\
\tilde{w}(p) = -\frac{4k_1}{(1+k_1)(a+2)p^2}
\eea
for the gauge propagator $\tilde G_{\mu\nu,\rho\sigma}(p)=\langle \tilde A_{\mu\nu}\tilde A_{\rho\sigma}\rangle$, 
\be
\tilde{f}(p)=\frac{2}{(1-2ka)p^2}\ ;\
\tilde{g}(p)=0\ ;\
\tilde{l}(p)=\frac{2}{(1+k_1)p^2}\ ,
\ee
for the propagator $\tilde G_{\mu\nu,\rho}(p)=\langle \tilde A_{\mu\nu}\tilde b_\rho\rangle (p)$ and, finally, 
\be
\tilde{r}(p)=\frac{4a}{2ka-1}\ ;\ \tilde{s}(p)=0\ ,
\ee
for the propagator $\tilde G_{\mu\nu} (p) = \langle \tilde b_\mu \tilde b_\nu\rangle (p)$. Hence, the gauge fixing condition \eqref{vectorgaugecond} leads to a set of propagators, as expected, which, as guessed, do not display anymore a pole in $a=0$, which was a signal of the change of symmetry of the action \eqref{Sinv}, and which still display poles in $a=1$ and $a=-2$, which have been discussed already. Notice the appeareance of another pole involving the $a$-parameter, at $2ka=1$, which should be unphysical, since it relates a physical parameter to a gauge unphysical one, together with an additional pole at $k_1=-1$, which occurs also in LG \cite{Gambuti:2020onb}. As a check, one can verify that the propagators computed here with the vector gauge condition \eqref{vectorgaugecond} coincide, at $a=0$ with those of massive LG at vanishing masses, as theey should \cite{Gambuti:2021meo}.

\subsection{Massive case}

From \eqref{massivematrixpropeq} we get
\bea
\tilde\Omega^{\mu\nu,\alpha\beta}(m_1,m_2)\tilde{G}_{\alpha\beta,\rho\sigma}(m_1,m_2) + 
\frac{p^2}{2}e^{\mu\nu}\tilde{G}_{\rho\sigma}(m_1,m_2) &=& {\cal I}^{\mu\nu}_{\ \ \rho\sigma} \label{6.6}\\
\tilde\Omega^{\mu\nu,\alpha\beta}(m_1,m_2)\tilde{G}^*_{\alpha\beta}(m_1,m_2)+\frac{p^2}{2}e^{\mu\nu}\tilde{G}(m_1,m_2) &=& 0 \label{6.7}\\
\frac{p^2}{2}e^{\alpha\beta}\tilde{G}_{\alpha\beta,\rho\sigma}(m_1,m_2) &=& 0 \label{6.8} \\
\frac{p^2}{2}e^{\alpha\beta}\tilde{G}^*_{\alpha\beta}(m_1,m_2) &=& 0\ .\label{6.9}
\eea
With the parametrization \eqref{massiveprop} we find from \eqref{6.8}
\be
\tilde{U}p_\rho p_\sigma + \tilde{W}\frac{p^2}{4}\eta_{\rho\sigma}=0\ \ 
\Rightarrow \tilde{U}=\tilde{W}=0\ ,
\label{}\ee
while from \eqref{6.9} we have
\be
\tilde{G}_{\mu\nu}(m_1,m_2)=2\frac{e_{\mu\nu}}{p^2}\ .
\label{}\ee
Therefore \eqref{6.7} becomes
\be
m_1\frac{e^{\alpha\beta}}{p^2}+\frac{m_2}{4p^2}\eta^{\alpha\beta}+p^2\tilde{G}(m_1,m_2)\frac{e^{\alpha\beta}}{2} =0\ ,
\label{}\ee
which implies
\be
m_2=0
\label{}\ee
and
\be
\tilde{G}(m_1,m_2)= -\frac{2m_1}{p^4}\ .
\label{}\ee
Taking all the above results into account, we can rewrite the condition \eqref{6.6} as
\be
[(t+m_1)A+m_1B+(v+m_1)C+(z+m_1)D]^{\alpha\beta,\mu\nu}
[\tilde{T}A+\tilde{V}C+\tilde{Z}D]_{\mu\nu,\rho\sigma}
+\frac{p^\alpha p^\beta}{2}\frac{2e_{\rho\sigma}}{p^2}
= {\cal I}^{\alpha\beta}_{\ \ \rho\sigma}
\label{}\ee
or
\be
[(t+m_1)\tilde{T}A +(v+m_1)\tilde{V}C+(z+m_1)\tilde{Z}D]^{\alpha\beta}_{\ \ \rho\sigma}
+e^{\alpha\beta}e_{\rho\sigma}=
[A+B+C+D]^{\alpha\beta}_{\ \ \rho\sigma}\ ,
\label{}\ee
so that
\bea
\tilde{T}(p;a,m_1) &=& \frac{1}{(2+a)p^2+m_1} \label{} \\
\tilde{V}(p;a,m_1) &=& \frac{1}{(a-1)p^2+m_1} \label{}\\
\tilde{Z}(p;a,m_1) &=& \frac{1}{ap^2/2+m_1}\ .\label{}
\eea

\end{document}